\documentclass[fleqn,usenatbib]{mnras}

\usepackage{newtxtext,newtxmath,comment}
\usepackage[T1]{fontenc}

\DeclareRobustCommand{\VAN}[3]{#2}
\let\VANthebibliography\thebibliography
\def\thebibliography{\DeclareRobustCommand{\VAN}[3]{##3}\VANthebibliography}

\usepackage{graphicx}	% Including figure files
\usepackage{amsmath}	% Advanced maths commands

\title[QPO amplitude and jet inclination]{The strength of Type-C quasi-periodic oscillations in black hole X-ray binaries correlates with the jet inclination}

\author[Vincentelli et al.]{
F. M. Vincentelli,$^{1,2}$\thanks{E-mail: federico.vincentelli@coventry.ac.uk}, 
N. Bollemeijer$^3$,
A. Veledina$^{4,5}$,
D. Altamirano$^2$,
Q. Bu$^6$,
F. Carotenuto$^7$,
P. Casella$^7$,\newauthor
Y. Cavecchi$^{8,9,10}$,
R. Ma$^2$,
G. Marcel$^4$,
G. Mastroserio$^{11}$,
S. Motta$^{12,13}$,
L. Zhang$^{14}$
\\
$^{1}$Centre for Fluid and Complex Systems, Coventry University,  CV1 5FB, Coventry, UK\\
$^{2}$School of Physics and Astronomy, University of Southampton, University Road, Southampton SO17 1BJ, UK\\
$^{3}$Anton Pannekoek Institute for Astronomy, University of Amsterdam, Science Park 904, 1098 XH Amsterdam, The Netherlands\\
$^{4}$Department of Physics and Astronomy, FI-20014 University of Turku, Finland\\
$^{5}$ Nordita, Stockholm University and KTH Royal Institute of Technology, Hannes Alfv\'ens v\"ag 12, SE-10691 Stockholm, Sweden\\
$^6$Institute of Astrophysics, Central China Normal University, Wuhan 430079, P.R. China\\
$^7$INAF-Osservatorio Astronomico di Roma, Via Frascati 33, I-00076, Monte Porzio Catone (RM), Italy \\
$^8$Departamento de Astrof\'isica, Universidad de La Laguna, 38206, San Crist\'obal de La Laguna, Tenerife, Spain\\
$^9$Instituto de Astrof\'{i}sica de Canarias, 38205, San Crist\'obal de La Laguna, Tenerife, Spain\\
$^{10}$Center for Nuclear Astrophysics across Messengers (CeNAM), 640 S Shaw Lane, East Lansing, MI 48824, USA\\
$^{11}$Scuola Universitaria Superiore IUSS Pavia, Palazzo del Broletto, piazza della Vittoria 15, I-27100 Pavia, Italy\\
$^{12}$Istituto Nazionale di Astrofisica, Osservatorio Astronomico di Brera, via E. Bianchi 46, 23807 Merate (LC), Italy\\
$^{13}$University of Oxford, Department of Physics, Astrophysics, Denys Wilkinson Building, Keble Road, OX1 3RH, Oxford, UK\\
$^{14}$Key Laboratory of Particle Astrophysics, Institute of High Energy Physics, Chinese Academy of Sciences, Beijing 100049, China\\
 }

\date{Accepted XXX. Received YYY; in original form ZZZ}

\pubyear{\the\year{}}

\begin{document}
\label{firstpage}
\pagerange{\pageref{firstpage}--\pageref{lastpage}}
\maketitle

\begin{abstract}
X-ray quasi-periodic oscillations (QPOs) are a characteristic feature of  low-mass X-ray binaries (LMXBs). These oscillations have been studied for decades and revealed a rich and complex phenomenology that is still not fully understood. \textit{RXTE} archival studies have shown that the amplitude of these oscillations differs significantly between black holes (BH) with high or low inclination. Yet, the actual dependence on inclination has never been adequately estimated. Thanks to the improvement of inclination measurements through radio observations and the recent observations by the \textit{HXMT} satellite, we quantified for the first time the dependence of {Type-C} QPO amplitudes on the jet inclination of individual BH LMXBs.  Our analysis reveals the presence of a significant linear correlation up to 8 Hz, strengthening the case for a ``geometrical''  origin of the QPOs.  In addition, for a given QPO frequency, we observe systematically lower amplitudes during the decay of outbursts compared to the rise.  This data collection represents a key benchmark for any QPO model.  Our comparison with the predictions from a precessing hot flow shows that the amplitude of the QPOs can be reproduced by this scenario if the spin-orbit misalignment is at least $\approx$10-15$^\circ$.
\end{abstract}

\begin{keywords}
keyword1 -- keyword2 -- keyword3
\end{keywords}

%%%%%%%%%%%%%%%%%%%%%%%%%%%%%%%%%%%%%%%%%%%%%%%%%%

%%%%%%%%%%%%%%%%% BODY OF PAPER %%%%%%%%%%%%%%%%%%

\section{Introduction}

Black hole low mass X-ray binaries (BH LMXBs) are galactic recur-
rent transients which contain a stellar-mass BH accreting matter from
a donor star with a mass of $\leq$1$M_\odot$ . During their active phase, these systems show a complex evolution in the X-rays, characterized by two
alternating main physical components  \citep{fender2004,dgk07}. In the initial phases of the outburst, referred to as the hard state,
the X-ray spectrum shows a strong non-thermal component, showing
a power law spectrum with an index which can vary between $\approx$1.5-2 \citep[see e.g.][]{zdariskigierlinski,Sobolewska2011}. This
is typically associated with an optically-thin, geometrically-thick ac-
cretion flow, sometimes referred to as the ‘corona’. As the outburst
continues, the spectrum becomes softer, and a thermal black body component appears. This is generally associated with an optically
thick, geometrically thin accretion disk \citep[see e.g.][]{shakurasunayev,kubota2004} . This latter regime is referred to as
the soft state. During the decay of the outburst, the spectrum remains
soft and goes back to a harder spectrum at a luminosity lower than the
hard-to-soft transition. The alternation between disc and corona dom-
inated spectra gives rise to a typical Q-shaped track when analysing
the evolution of X-ray hardness versus intensity\citep{belloni2005}.
The origin of this behaviour and the exact accretion/ejection geometry evolution during the outburts are still poorly understood, and still
a matter of debate \citep[see e.g.][]{marcel2020a,demarco2021,wang2022,Kawamura2022,uttleymalzac}.

BH LMXBs are also known to show strong emission at lower energies, from radio to O-IR wavelength \citep{corbel2002,gandhi2011}. Especially during the hard state, a flat spectral component is observed across this wavelength regime. While the radio emission  is typically explained in terms of synchrotron emission from a compact jet \citep{blandford&konigl,peer2014}, it has been shown that the O-IR band can also have a contribution from the hot flow via synchrotron emission \citep{veledina2013,veledina2017} or irradiation \citep[e.g.][]{hynes2009}. Intensive monitoring of these objects showed that while the jet is quenched in the soft state, strong discrete ejecta are observed during the hard-to-soft transition \citep{fender2004,bright2020}. Tracking of the proper motion of these blobs has allowed us to estimate key quantities of the jet, such as inclination, energetics and properties of the surrounding medium \citep{mirabel1994,bright2020,carotenuto2022,carotenuto2024}.

One of the most remarkable properties of BH LMXBs  is the presence of strong, stochastic fast variability in their X-ray emission \citep{vanderklis1989,belloni2000}. Such fluctuations are mainly observed in the hard state and are associated with the presence of the hot Comptonizing medium. In the Fourier domain, this variability shows a very broad, band-limited power density spectrum  (PDS) that spans over a wide range of frequencies (i.e. between $\approx$ mHz and $\approx$100 Hz). Characterization of this variability led to the identification of characteristic breaks in the PDS, which evolve during the outburst, showing a consistent shift towards higher frequencies as the system moves towards the soft state \citep{belloni2005}.

 Since the early X-ray timing studies, it has also been known that LMXBs show narrow quasi-periodic oscillations \citep[QPOs; see e.g.][]{belloni2002,casella2004} between 0.01 and 100 Hz. In BH LMXBs, QPOs can be classified as  High frequency, if found above $\approx$60 Hz, or  low frequency, if found below $\approx$30 Hz \citep{ingram_notta2020}. Low frequency QPOs have been divided in three categories \citep{{casella2004}}: \textit{Type A} are weak signals, which are observed during the soft state at  $\approx$8 Hz and have been rarely detected; \textit{Type B} have been almost exclusively observed in the latest stages of the transition between the hard and soft state. They are distinguishable by the absence of broadband noise and are typically observed in the 4--8 Hz range \citep{motta2015}; finally, \textit{Type C} are typically seen across the hard and hard-intermediate state in the 0.01-10 Hz range,  and can be identified by the presence of broadband noise. Type C QPOs also evolve during the outburst, and as the characteristic frequencies of the noise, they also shift towards higher frequencies as the system moves towards the soft state. Given the large range in frequency and their presence across the most luminous stages  of the hard/hard-intermediate state, historically, most of the efforts on explaining low-frequency QPOs have focused on Type C. In this paper, we will focus only on this QPO class.

Despite decades of studies,  the physical origin of Type C QPOs remains uncertain. Since the early 90s, the coherence of these oscillations has led  to associating this phenomenon with some kind of Keplerian motion close to the central object \citep{Alpar1992,miller1998,stella1998}.   This link was strongly supported by the presence of a significant correlation between the frequencies of the QPOs and the breaks in broadband noise \citep{psaltis1999}. Such a correlation was successfully modelled with the relativistic precession of rings of matter, allowing to put tight constraints on the fundamental parameters of the central BH: i.e. mass and spin \citep{stella1999,motta_2014_1655,motta2014_1550}.   

Although some models also attempted to reproduce the QPOs through some kind of accretion instability (see e.g. \citealt{tagger1999,varnier2002,ingram_notta2020}), further observational studies consolidated the geometrical origin. One of the most constraining results has been the discovery of a dichotomy in the QPO amplitude between high inclination and low inclination sources, with the first showing systematically higher amplitude than the second \citep{Schnittman2006,motta2015}. Later works also showed that other QPO properties also depend on inclination: the phase lag of the fundamental \citep{vandeneijnden2017} and the waveform \citep{deRuiter2019}.  These results strongly supported models based on the precession of the inner hot flow \citep{ingram2009,ingram2012,zycki2016}. In particular, follow-up studies showed that a truncated geometrically thick disc undergoing Lense-Thirring precession could explain the frequency evolution of the power spectrum through the outburst, as well as its spectral modulation \citep{ingram2010,ingram2016,ingram2017,you-bursa-zycki2018}.

Despite the success of this interpretation, new studies showed an inconsistency between the hot flow's external radius inferred from Lense-Thirring precession and the one obtained with other spectral-timing observations \citep[see e.g.][]{marcel-neilsen2021,nathan2022}. More recently, alternative approaches have attempted to model QPOs also by fitting their X-ray rms and lag spectrum with a Comptonization model from a spherical cloud, without any assumption on the origin of the oscillations  \citep{bellavita2022,mendez2022}. Although this approach is still partly phenomenological, it did successfully reproduce the evolution of several BH LMXBs with one or two Comptonizing clouds of variable size \citep[see also][]{ma2023,alabarta2025}.

Given the uncertainty regarding the origin of the QPOs, it is important to build a general test for any model that wants to reproduce this phenomenon.
For instance, the spectral-timing differences between the rising and decaying phases of the outbursts offer a strong opportunity to shed light on the origin of QPOs. \citep{kalemci2006,munoz-darias2010,vandat2019}. It has been shown that the overall broadband absolute variability is lower during the decay compared to the rise \citep{munoz-darias2010}. Moreover, recently it has been shown that, at the same frequency, the  QPO rms of H 1743--322 during the decay is significantly lower than during the rise \citep{shui2023}. Despite this, a detailed comparison of the QPO properties between these two states is still missing.

Also, as shown by past studies, the inclination dependence of QPOs is clearly a very powerful diagnostic \citep{Schnittman2006,motta2015}. Yet, the actual trend has never been adequately quantified so far.  \citet{Schnittman2006}  did attempt to quantify it with direct inclination measurements, but their sample was too small to reach strong conclusions.  Given this, later works divided sources into only broad categories (i.e. ``high'' and ``low'' inclination systems; see also \citealt{vandeneijnden2017}). 

This paper aims to address the current gap and to re-perform a systematic study to determine the actual dependence of the amplitude on inclination.  It is important to highlight that different kinds of inclination can be measured in LMXBs: one associated with the orbital plane, one with the accretion flow, and another with the direction of the jet. While it is usually assumed that the jet is perpendicular to the disc, both theory and observations do indicate that these two components are not necessarily orthogonal \citep{fragos2010,liska2018,miller-jones2019,Poutanen2022,Zdziarski2023}. This suggests caution when selecting the inclination of the system. Given this, we decided to study only objects that have well-defined jet inclinations.

\section{The sample} 
\label{sec:data}

With this work we want to connect three quantities: \textit{ jet inclination, QPO amplitude} and \textit{QPO frequency}.    For the jet inclination, we considered the measurements obtained from the ballistic motion of the discrete ejecta during the hard-to-soft spectral transition. A recent study by \citealt{fendermotta2025} has shown that LMXBs have two different kinds of jets: fast ones ($v_{\rm jet}/c = \beta>0.9$), which have a fixed direction and slow ($\beta<0.9$) ones, which can change orientation. Our selection focuses only on the fast ones.  Regarding QPOs,  given their typically hard spectrum \citep{casella2004,casella2005}, we focused our search on the sources detected by instruments with an effective area between 2 and 25 keV: \textit{RXTE} and \textit{HXMT}. %In particular, we measured the same energy range used in the previous sample study with RXTE by \citet{motta2015}.

After a literature search, we found six BH LMXBs which met our criteria in the \textit{RXTE} archive: 4U~1543--47, XTE~J1550--564, GRO~J1655--40,  H~1743--322, XTE~J1752--223 and GRS 1915+105. 
Within the \textit{HXMT} archive, we considered  three very bright recent transients: MAXI J1348--630, MAXI J1820+070 and Swift J1727.8--1614. While only the first two have well-defined jet-inclination measurements, we decided to also include the third object due to the exceptional wealth of QPO detections.  {A summary of the objects used for this work, including mass and inclination of each system, is shown in Tab. \ref{taB:tab}.  Not all systems contain a dynamically-confirmed BH. Among the ones with actual measurements, the masses span a range from $\approx$5M{$_\odot$} to  $\approx$12 M{$_\odot$} \citep{Corral-Santana2016}.} 

\begin{table*}
\caption{Average \textit{rms} in the 2--25 keV band for all the sources in our sample during their rising phase. QPOs were divided into four different frequency bins. For each source, we also list the satellite used for the rms measurement, the jet inclination angle ($i_{\rm jet}$), {the BH mass (when known) and the references from which these measurements were taken}. The number in parentheses indicates the number of QPOs used to compute the average {rms} for that frequency bin.}
\label{taB:tab}
\begin{tabular}{llcc|ccccccl}
\hline
&&&&&&\\
        &  & && & & \textit{rms} (\%) & & \\  
Satellite &         Source &  $i_{\rm jet}$  ($^\circ$)& BH Mass ($M_\odot$)& <1 Hz &  1--4 Hz& &  4--8 Hz & 8--10 Hz & Ref. \\
\hline
 &&&&&&\\
\textbf{RXTE} &&&&&&\\

 &4U 1543--47     & <30 & 8-10& --  & -- & &  2.1$\pm$0.3 (3)          &  --              &   [1,2]      \\     
&XTE J1550--564  &  73.5$\pm$9   &8-14 &    14.5$\pm$0.2  (20)  &   13.1 $\pm$ 0.2 (20)   & & 9.5$\pm$ 0.6 (25)             &      --        &       [3,4] \\

&GRO J1655--40 &   85$\pm$2   & 6$\pm$0.4& 11.3$\pm$1  (10) &    21$\pm$0.3 (4)    && --              &  --  &   [5,6]    \\

&H 1743--322      &  75$\pm$5&--& 14.5$\pm$0.4 (14)        &13.8 $\pm$0.4 (23)&&          8.4$\pm$0.6   (13)    &       4.6$\pm$0.1  (1)            &        [7]    \\
&XTE J1752--223    & 18$\pm$6&--&  --       & 3.5$\pm$0.01(1) &&     4.1$\pm$0.2 (1)           &     --          &   [8]         \\

 & GRS 1915+105 &60$\pm$3&10-12& 10.4$\pm$0.9 (52) & 12.7$\pm$1.3 (464)& & 7.5$\pm$1.8 (104) & -- & [9,10]\\
&&&&&&\\
\textbf{HXMT} &&&&&&\\
 
&MAXI J1348--630    & 30$_{-10}^{+20}$    &--  & 6.5$\pm$0.1 (7) & 5.5$\pm$0.3 (1)  &  & 4.2$\pm$0.1 (2)&-- & [11] \\ 
&MAXI J1820+070    &  60$\pm$1 &8$\pm$1& 10.7 $\pm$0.4  (8)    & -- &  & -- &-- &[12,13] \\
&Swift J1727.8--1613  &   <67 &  >3&14.4$\pm$0.2 (28) &  12.5$\pm$0.1 (128) && 6.4$\pm$0.2 (62)&  2.9$\pm$0.1 (3)& [14,15]  \\
\hline

\end{tabular}
\\[6pt]
\textbf{References:} [1] \citet{ZhangZ_2025} [2] \citet{Orosz1998} 
[3]\citet{steiner2012} [4] \citet{Orosz2011}
[5]\citet{hjelleing1995} [6] \citet{Shahbaz2003}
[7]\citet{steiner2012_1743} [8] \citet{carotenuto2024} [9]~\citet{ried2023} [10] \citet{vincentelli2023nat} [11] \citet{carotenuto2022} [12]~\citet{cooper2025}  [13]\citet{torres2020} [14] \citet{wood2025} [15]\citet{matasanchez2025}\\

\label{tab:all}
\end{table*}

\section{Analysis}
\label{sec:analysis}

\subsection{Quantifying the QPO amplitude}

\textit{RXTE:} Systematic studies of the QPO of the \textit{RXTE} archive have been performed by \citet{motta2015} and \citet{ZhangL}. To be consistent with previous studies, we estimated QPO properties using the same methodology and energy range used by \citet{motta2015} (2--25 keV).  Power spectra were computed using 128-s-long segments up to a Nyquist frequency of 1024 Hz. Observations were fitted using a multi-Lorentzian model. We used observations which had low-frequency ($<$30 Hz), relatively narrow features (quality factor Q$>$ 2) on top of the broadband noise. Only QPOs detected at a significance greater than $3\sigma$ were considered. Unlike \citet{motta2015}, however, we focused only on the \textit{amplitude of the fundamental peak} (i.e. the strongest peak in the power spectrum).  The rms was computed as the area of the Lorentzian associated with the QPO.

\textit{HXMT:}  None of the three instruments of \textit{HXMT} (LE, ME and HE) entirely share the same energy range of the \textit{PCA} on \textit{RXTE}. Thus, to compare the results with \textit{RXTE}, we combined the QPO amplitudes measured in the  2--10 keV (\textit{LE}) and 10--25 keV (\textit{ME)} bands to obtain an equivalent 2--25 keV rms. The fractional squared rms can be obtained using the following weighted average (see Appendix for the formal derivation):

  \begin{equation}
  \begin{split}
rms_{equiv.}^2 = rms^2_{LE} \cdot W_{LE}^2 + rms^2_{ME}\cdot W_{ME}^2 \\
+ 2 \sqrt{rms_{LE}^2 \cdot W_{LE}^2 \cdot rms_{ME}^2 \cdot W_{ME}^2}
\end{split}
\label{eq:rmseq}
  \end{equation}
where the weights for each instrument $W_{XX}$ are defined as:
\begin{equation}
\label{eq:weight}
W_{XX}^2=\frac{<CR_{XX}>^2}{<CR_{LE}+ CR_{ME}>^2} 
\end{equation}
and $CR_{XX}$ represents the count rate in a given instrument ($LE$ or $ME$).

\begin{figure}
    \centering
    \includegraphics[width=\linewidth]{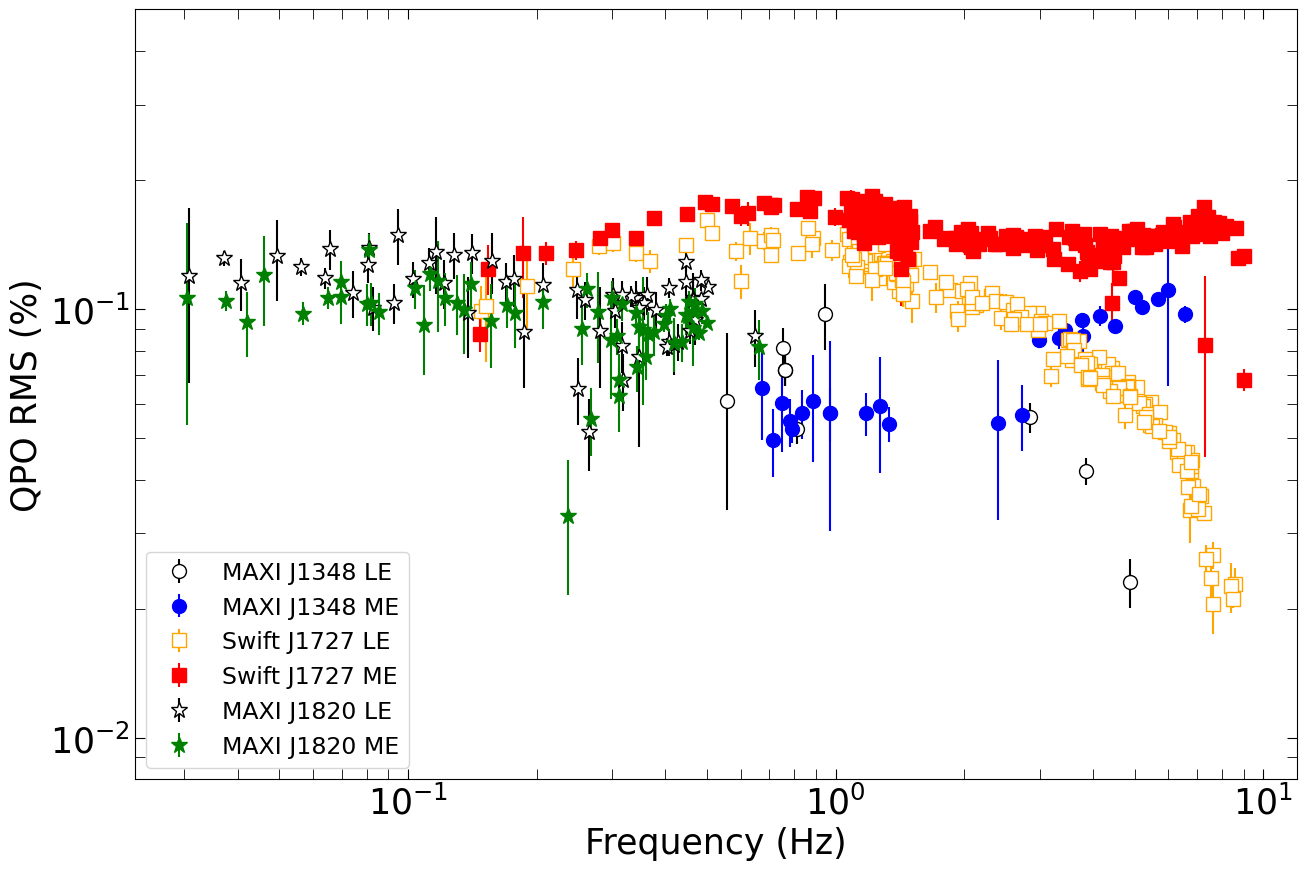}
    \caption{The individual fractional rms measurements of the QPO for three sources with the \textit{HXMT} \textit{LE} and \textit{ME} instruments in the 2--10 and 10--25 keV range. At low QPO frequencies, the measured rms values are similar, while above ${\sim}2$ Hz, the QPO is much stronger in the harder energy band.}
    \label{fig:HXMT_LE_ME}
\end{figure}

 We proceeded by extracting the background-subtracted and deadtime-corrected light curve files (\texttt{lcnet}) using the \texttt{hpipeline} command of \texttt{HXMTDAS} v2.06 for \textit{LE} and \textit{ME} in the 2--10 and 10--25 keV energy range. Using those light curve files, we computed the PDS in fractional units following \citet{vaughan2003}. We used a 1/512 s time resolution and estimated the Poisson noise level, which can be affected by deadtime, by fitting a constant to the power spectrum in the 200--256 Hz range, where the intrinsic source variability is negligible. We then obtained the rms by fitting the PDS with a multiple Lorentzian fit \citep{belloni2002}. The resulting rms vs frequency plots of the LE and ME are shown in Fig. \ref{fig:HXMT_LE_ME}. At low frequencies, the QPO rms is similar in both energy bands, although for MAXI J1820+070, the rms is slightly higher between 2--10 keV and vice versa for Swift J1727.8--1613. At higher QPO frequencies, the LE QPO rms decreases while the ME rms remains high for both Swift J1727.8--1613 and MAXI J1348--630. As also shown in the plot, some of MAXI J1348--630 observations showed a significant QPO only in the ME band: these points were excluded from the analysis.

\subsection{QPOs during the outburst  decay}

As noted in \citet{motta2015}, even when considering a single object, the rms-frequency relation can show a very large scatter. This represents an obstacle to quantifying the actual dependence on inclination. Interestingly,  \citet{shui2023} recently showed that the rms of the high-inclination source H 1743--322 is significantly lower in the decay than in the rise of the outburst. Thus, as a first step, we decided to investigate further this effect and inspected the individual relation of the objects with high number of observations and a clear transition back to the hard state: XTE~J1550--564, GRO~J1655--40 and H 1743--322.

\begin{figure}
\centering
\includegraphics[scale=0.4]{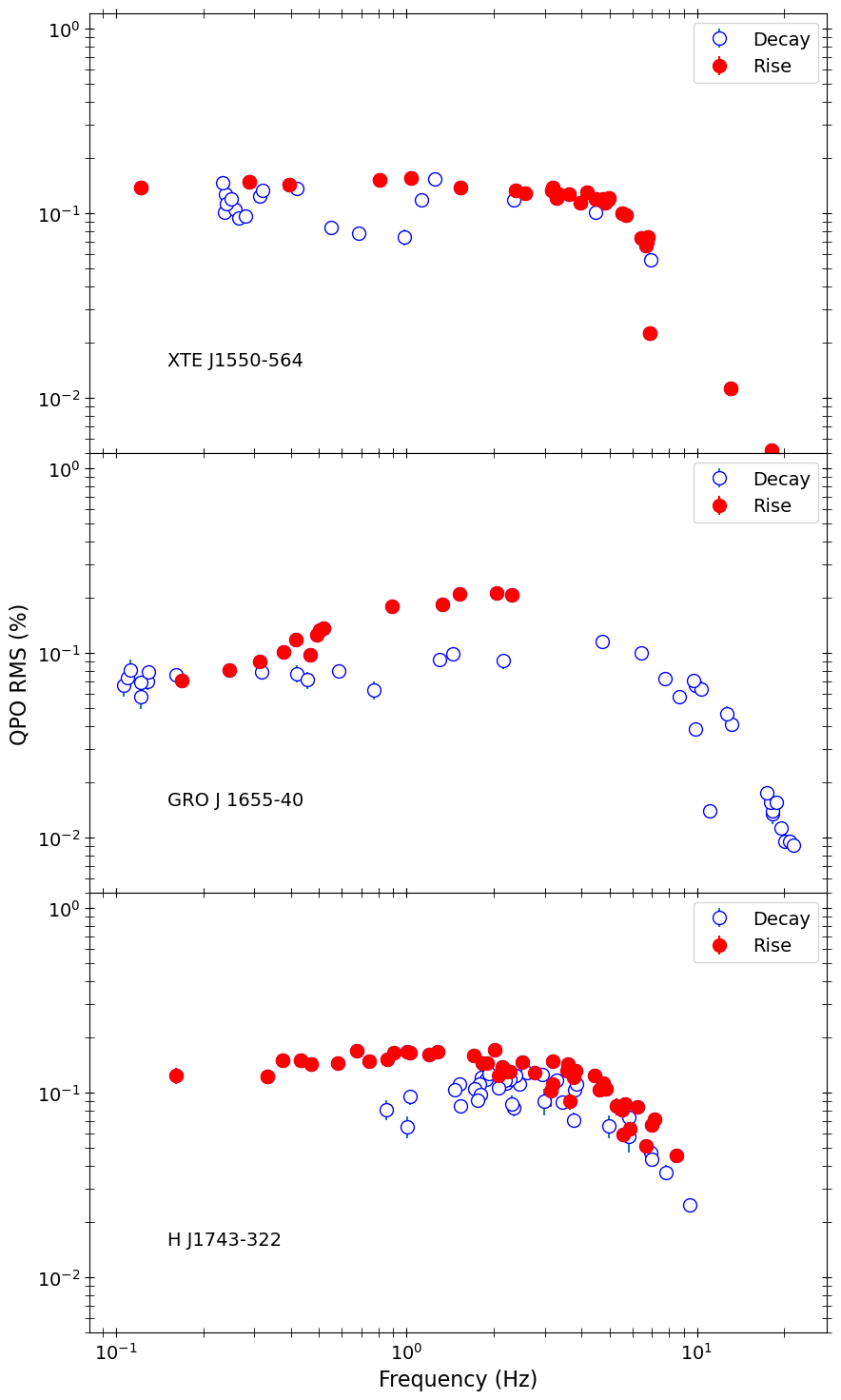}
\caption{RMS vs frequency diagram for the 3 sources  with the highest inclination and the highest number of detections. The red points represent the data in the rising phase of the outburst, blue points were observed during the decay.\label{fig:decay}}
\end{figure}

We divided the observations into two categories: those taken during the rising phase of the outburst (i.e. before the transition to the soft state) and those taken during the decay (i.e. after the reverse transition). The results for the three objects are shown in Fig. \ref{fig:decay}. All objects show that the decaying phase has a lower QPO amplitude than the rise, especially in the range between $\approx$0.5 and $\approx$2--3 Hz. Ignoring for the moment the physical reasons behind such a behaviour, this  clearly shows that the QPOs from these two states have to be analysed separately. Thus, from this point onwards, only the QPOs from the rising part of the outburst will be considered.

\subsection{QPO evolution vs frequency and inclination}

\begin{figure}
\centering
\includegraphics[width=\linewidth]{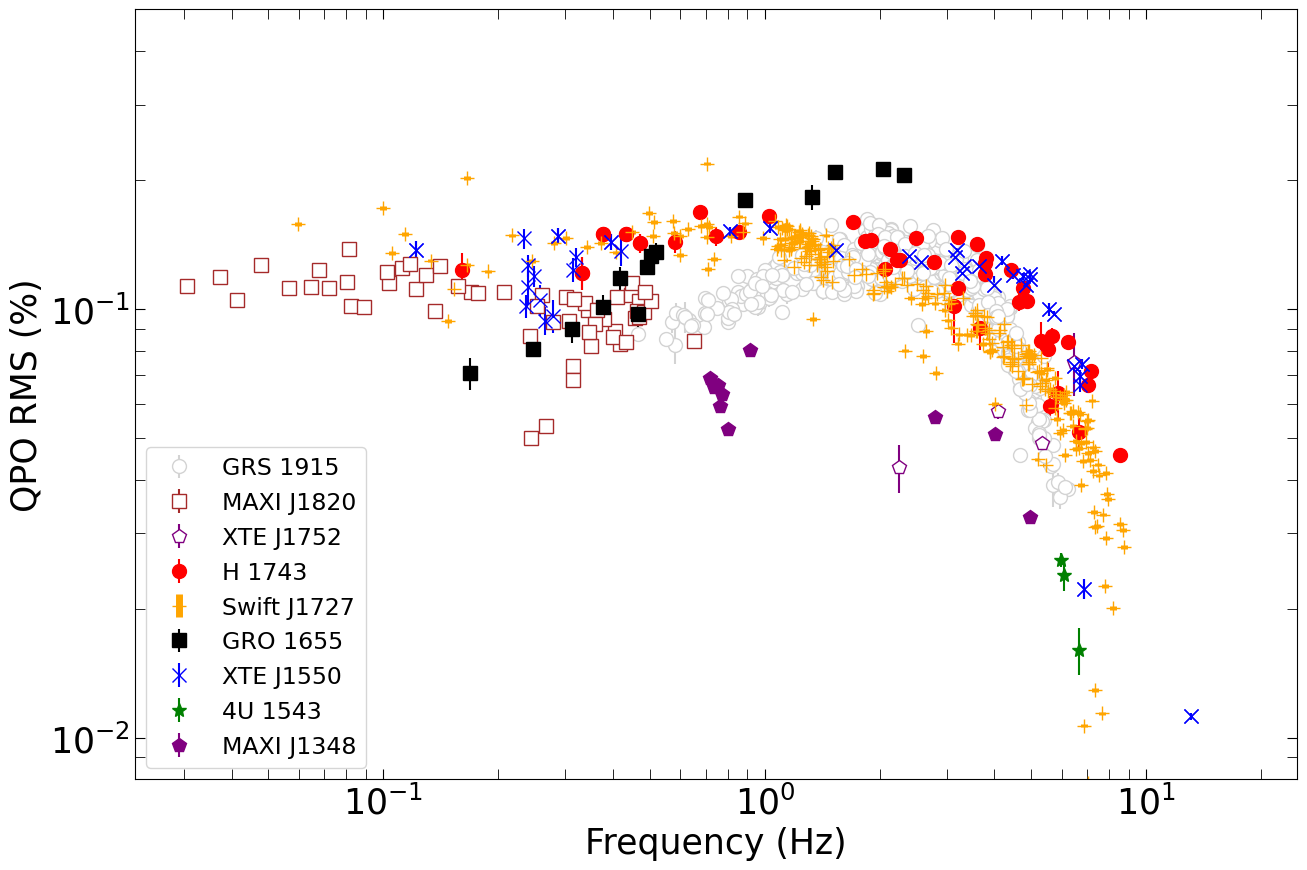}
\caption{2--25 keV average RMS vs frequency diagram of our sample using data only from the rise of the outburst. Different sources, each with a different jet inclination, have clear individual tracks.
\label{fig:rmsfreq}}
\end{figure}

In Fig. \ref{fig:rmsfreq} we show the resulting rms-frequency diagram using the sources from both our \textit{RXTE} and \textit{HXMT} samples. 
By checking the evolution of each object individually, it is clear that different sources draw specific tracks, defining a somewhat specific  \lq\lq plateau\rq\rq~ for $\nu_{\rm QPO}<4$\,Hz, before showing a significant drop in rms for $\nu_{\rm QPO}>4$\,Hz.  Given this, similarly to \citet{Schnittman2006}, we  estimated the average rms for 4 different QPO frequency ranges:  0--1, 1--4, 4--8 and 8--10 Hz. If more than 3 data points per bin were found,  we computed the errors in each bin as the standard deviation, otherwise standard propagation was used.

\begin{figure*}
\centering
\includegraphics[scale=0.45]{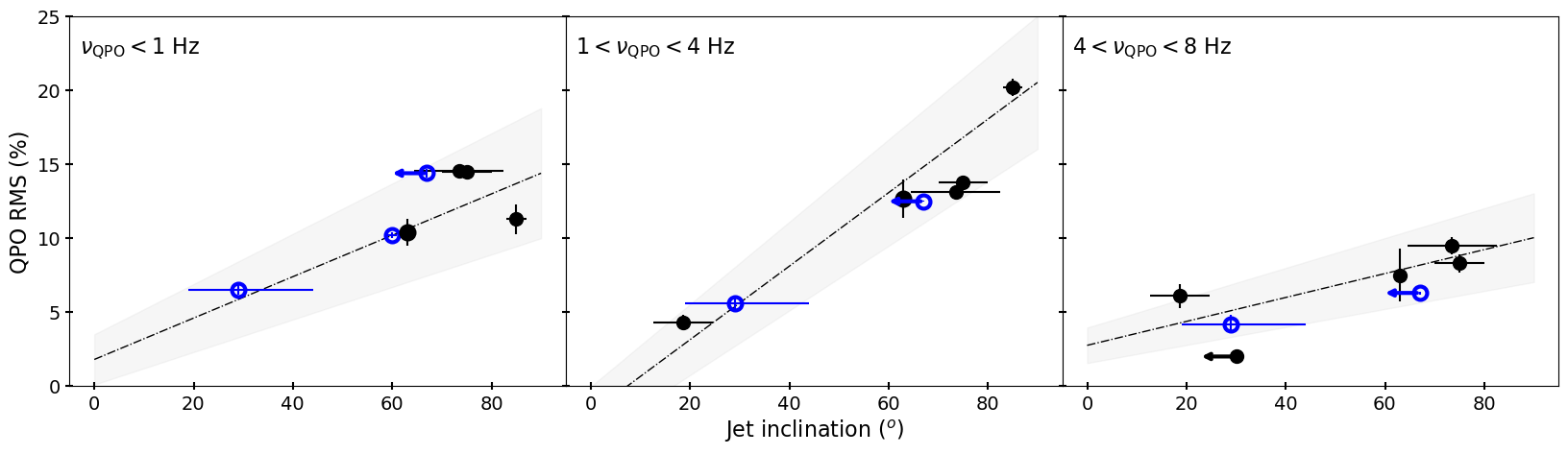} 
\caption{2--25 keV QPO RMS vs jet inclination in three frequency ranges:  $<1$~Hz (Left panel), 1--4 Hz (Central panel) and 4--8 Hz (Right panel). Black-filled points are obtained from the \textit{RXTE} archive, while empty blue points are obtained from \textit{HXMT}. All frequency bands exhibit significant correlations. The dashed lines represent the best linear fits using an orthogonal distance regression, and the grey areas represent the 1 $\sigma$ confidence intervals.}
\label{fig:rms-inc}
\end{figure*}

We then tested whether there was any trend between the QPO amplitude and jet inclinations (see Tab. \ref{taB:tab}). Given that the 8--10 Hz band had only two sources, we discarded those points from the rest of the analysis. The resulting rms-inclination diagrams are shown in Fig.~\ref{fig:rms-inc}.  Not all objects have a detection in all frequency ranges; thus, not all diagrams have the same number of data points.   Our results suggest the existence of a linear correlation between the two quantities, but, due to the limited number of sources, other trends are still allowed (see also Sec. \ref{disc:precession}). In general, the correlation appears to become shallower above 4 Hz.

Given the low number of datapoints we quantified the correlation between rms and inclination using both Pearson and Spearman coefficient (performed excluding points with only upper limits on the inclination). In both cases, the tests  revealed a high correlation coefficient across all bands (see Tab. \ref{taB:fit}). The tightest correlation was observed between 1 and 4 Hz, while the lowest was below 1 Hz.  We also performed a linear fit on the data to quantify the actual trend. Given the large errors in the inclination,  we used the \textsc{scipy} implementation of the Orthogonal distance regression\footnote{\url{https://docs.scipy.org/doc/scipy/reference/odr.html}}, which accounts for errors in both the x and y axes \citep{BoggsDonaldson1989NISTIR,SciPy2020}. Also in this case, upper limits were excluded from the fit. The results are reported in Tab. \ref{taB:fit}, and the resulting curves are shown in Fig. \ref{fig:rms-inc}. As reflected in the correlation coefficient, the tightest relation is between 1 and 4 Hz.

 \begin{table*}
\caption{Correlation analysis. We report the number of data points used, the Pearson and Spearman correlation coefficient $\rho$, its associated $p$-value and the results of the linear fit (${\rm rms}=m\, i_{\rm jet}+q$) of  the rms vs inclination data in the different frequency bands. The last column also shows the sum of the squared residuals (on both the x and y axes) divided by the number of degrees of freedom.}
\label{taB:fit}
\begin{tabular}{l|ccccccr}
\hline
Band & $N_{\rm points}$ & $\rho_{\rm P}$ ($p$-value) &  $\rho_{\rm S}$ ($p$-value) & $m$ & $q$ & S/d.o.f. \\
\hline
\\
 <1Hz & 6 & 0.81 (0.04) & 0.77 (0.05) &  0.14 $\pm$0.03 & 1.9$\pm$1.7  & 4.2 \\
 \\
1--4 Hz & 6 &0.96 (0.003)&  1 (0.001) &  0.25$\pm$0.03  &   $-$1.8 $\pm$ 1.8 & 2.2 \\
\\
4--8 Hz & 5 &0.86 (0.06) & 0.79 (0.05) &  0.08 $\pm$0.02 &  2.7$\pm$ 1.3 & 2.1 \\

\hline
\end{tabular}

\end{table*}

\section{Discussion}
\label{sec:discussion}

We performed a systematic analysis of the dependence of Type C QPO amplitude on jet inclination, dividing \textit{RXTE} and \textit{HXMT} data into various frequency bands. Our analysis showed two  main results:

1) For sources with an inclination higher than $\approx70^\circ$, the amplitude of the QPOs during the decay is significantly lower than during the rise.

2)  The amplitude of Type C QPOs is tightly correlated with the jet inclination.

QPOs are fundamental features of accreting BH LMXBs, and yet they are still very poorly understood. Both these results have implications on our understanding of this key phenomenon. In the next  sections we will go through these points individually.

\subsection{On the different rms between rise and decay}

Our analysis reveals that, at comparable QPO frequencies, the fractional rms during the decay phase is systematically lower than during the rise, particularly in high inclination systems. This could explain why no QPOs are observed during the decay of most BH LMXBs. Given that low-inclination systems already exhibit intrinsically lower amplitudes, the additional attenuation observed during the decay makes the detection during this phase very challenging. In this regard, it is also interesting to notice that there has been growing evidence of weaker QPOs during the decay also in lower inclination systems. On one hand, evolving optical-infrared QPOs have been detected during this state, without a significant X-ray counterpart \citep{gandhi2010,vincentelli2019}. On the other hand, recent analyses of \textsc{nicer} data are revealing the presence of so-called ``hidden'' components, which are visible only in the cross-spectrum and not in the power spectrum \citep{alabarta2025,bellavita2025}. Although the interpretation of these ``imaginary'' QPOs  is still under discussion \citep[see also][]{veledina_poutanen_ingram_2013,veledina2018,konig2024_cyg,fogantini2025}, their lower amplitude QPOs are in line with our findings. 

In this regard, it is interesting to notice that during the outburst of MAXI J1820+070, the source luminosity decreased by a factor of a few during the bright hard state, called the ``bright decline'' by \citet{demarco2021}. During the bright decline, the QPO amplitude decreases, as is visible in Figs. \ref{fig:HXMT_LE_ME} and \ref{fig:rmsfreq} as the track between $\sim$0.2 and 0.4 Hz, where the QPO amplitude decreases for lower frequencies (and luminosities). This may indicate that the discrepancy in QPO amplitude between the rise and decay is caused by their different accretion rates and not by the phase (rising/decaying) of the source (see also \citealt{Bollemeijer2025b}). 

From a physical point of view, the presence of weaker QPOs and broadband noise in the lower branch of the hardness intensity diagram indicates that the accretion flow has properties different from those of the rising phase. Different kinds of observations are supporting this scenario. For the case of H 1743--322, \citet{shui2023} showed that the different rms can be explained with a different aspect ratio of the hot flow.    Interestingly, different radiative properties between the rise and decay of the outburst were also found by \citet{barnier2022} using X-ray and radio measurements of GX 339--4.  More recently, IXPE observations of Swift J1727.8--1613 showed that the same polarization degree and angle of the rise are recovered back in the hard state decay, suggesting a similar geometry \citep{Veledina2023,Ingram2024,Podgorny2024} This was well modeled by using a medium that, compared to the rising phase, has a higher cut-off temperature and a lower optical depth \citep{Podgorny2024}.

\subsection{On the rms vs jet inclination correlation}

The inclination of the system is expected to play a key role in shaping the phenomenology of BH LMXBs \citep{Schnittman2006,munozdarias2013,heil2015,motta2015,motta2018,reig2019}. Due to the difficulty in measuring the inclination, most of the studies have focused on dividing the sources into two broadly-identified high and low inclination groups. Thanks to the recent advancement in radio facilities and the significant increase of QPO detections with HMXT, in this work, we used measurements from individual sources, allowing us to quantify the actual rms vs jet inclination dependence.  Our results show that a significant correlation is clearly present across a wide range in frequency, confirming that inclination is a key factor for the detectability of QPOs. This result strongly favours a non-axisymmetric, geometrical modulation mechanism. Any model invoking purely isotropic luminosity oscillations in a spherical configuration is naturally disfavoured by the observed inclination dependence.

We also notice that above 4 Hz, this correlation is weaker than at lower frequencies. This is most probably due to a strong disc-component contamination during the hard-to-soft spectral transition. This might also explain why the rms-frequency evolution of the sources shows a hard cutoff around 4--8 Hz. This can be clearly seen in the rms evolution of Swift J1727.8--1613, which shows a drop only below 10 keV (Figure \ref{fig:rmsfreq}). A detailed analysis (beyond the scope of this paper) of the \textit{absolute} rms spectrum of these QPOs is necessary to adequately disentangle the intrinsic variability of this component {\citep[see e.g.][]{you-bursa-zycki2018}. On this regards it is also interesting to  note that the  the cut-off frequency of the rms seems to have a spread around 5--10 Hz. A possible origin of this behaviour might be due to the different masses of the sources. We can expect a scaling law with mass for the QPO properties from both observational \citep{motta_2014_1655,motta2014_1550} and theoretical arguments \citep{ingram_notta2020}.  The systems of our sample that have a well defined mass and high frequency cut-off are only GRS 1915+105 ($\approx10-12 M_\odot$; \citealt{Reid2014,ried2023,steeghs2013} ), XTE J1550--65O; ($M_{\rm BH}\approx8-14 M_\odot$ \citealt{Orosz2011}) and (if we consider also data from the decay) GRO 1655--40; ($\approx6 M_\odot$; \citealt{Shahbaz2003}). This limitation does not allow us to adequately investigate the dependence of this effect on mass. In the future, a larger sample of sources could allow us to understand how the QPO properties in general depend on mass (and/or other parameters).}

As mentioned above, several models have been developed to explain QPOs. Our new rms-inclination dataset, therefore, can place strong constraints on the different proposed scenarios (for reviews and discussion about these models, we refer the reader to \citealt{ingram_notta2020,ferreira2022,buisson2025})

One of the earliest models proposed to explain low-frequency QPOs is the accretion–ejection instability, in which a global spiral density wave forms in the inner regions of a magnetised accretion disc and can modulate the X-ray emission \citep[see e.g.][]{tagger1999}.  While it has been shown by \citet{motta2015} that these kinds of models cannot explain all the different types of QPOs, the non-axisymmetric structure which arises in this instability is such that an inclination dependence of the QPO is expected \citep[see e.g.][]{varniervincent2016,varniervincent2017}. However, according to the presented calculations, the intrinsic amplitude of the oscillation should increase with frequency. This does not seem to be the case for a large fraction of sources,  even above 10 keV, where the disc contribution is negligible (see e.g. Figure \ref{fig:HXMT_LE_ME} or also \citealt{li2013,ZhangL}). Further tests of this scenario are therefore needed to assess whether different model parameters could reproduce the observed rms-inclination QPO relation.

A model that has recently become popular relates how the hot Comptonizing medium would respond to an external perturbation \citep[see e.g.][]{karoupzas2020,garcia2021,mendez2022,Mastichiadis2022,ma2023,alabarta2025,zhang2026}. By assuming a spherical geometry \citep[see, e.g.,][]{bellavita2022}, this approach successfully reproduced the observed energy, rms and lag spectrum, constraining the size of the Comptonizing medium.  Similar radiative models have been proposed in the past: for example, \citealt{Cabanac2010} have also shown that  magneto-acoustic perturbations traveling from the disc to a slab corona could reproduce the QPOs. In such radiative models, based on small perturbations, the projected emitting area remains independent of viewing angle. As a consequence, the observed modulation amplitude should not depend on inclination. Reconciling such models with the observed inclination dependence of the QPO properties would therefore require introducing additional structural modulations or geometric asymmetries. A possible example of how to reconcile such models with the data could be considering vertical oscillating modes in a slab corona. This would produce stronger oscillations when viewed at high inclinations\footnote{Radial oscillation modes, instead, will be stronger when viewed face on.}. We notice that a similar scenario has been envisioned by \citealt{ferreira2022}, who discussed the possibility of an instability in the jet traveling to the hot flow. However, further modeling is still required to test whether these modes can actually exist and if they can reproduce the observed amplitude.

A third kind of scenario is based on the orbital motion of a portion of the inner disc.  One of the most successful models, usually referred to as the ``relativistic precession model'' (RPM), was able to explain the evolution in frequency of the QPO and the other variable components by linking them to the Keplerian, nodal and epicyclic frequencies of a test mass orbiting around the central star \citep{stella1998,stella1999,motta_2014_1655,motta2014_1550}. Yet, we note that the model, as it currently stands, assumes small perturbations only and does not provide quantitative predictions for the amplitude. Thus, it cannot be tested with the rms-inclination QPO relation.

The success of the RPM in reproducing the observed QPO behaviour rapidly led to the  development of models based on Lense-Thirring (LT) precession of the inner hot flow \citep{fragile2007,ingram2009}. The finding that QPOs are stronger in higher-inclination systems \citep{Schnittman2006,motta2015} naturally supported this framework \citep{veledina_poutanen_ingram_2013} and led to detailed modeling of the spectral modulations in these systems \citep{ingram2016,ingram2017},{ including their inclination dependence \citep{you-bursa-zycki2018,You2020}. A key element that emerged from this approach is the strong dependence of the rms spectrum on the azimuthal angle between the black hole spin and the normal to the  orbital plane \citep{veledina_poutanen_ingram_2013,you-bursa-zycki2018,You2020}. Given that such an angle has been measured only in one source \citep{Poutanen2022}, this made these models difficult to test. By using the jet inclination,} our compilation of data represents a key step forward in testing this scenario. While a detailed, source-by-source modeling of this dataset is beyond the scope of this work, in the next section, we show how computing basic predictions from  precession provides valuable physical insights into QPOs.

\subsection{Precession of a Comptonizing slab}
\label{disc:precession}

\begin{figure}
\centering
\includegraphics[width=\linewidth]{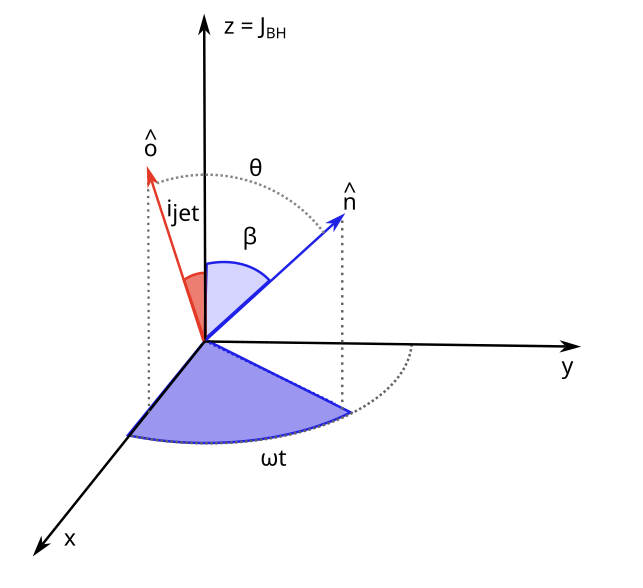}
\caption{{Schematic illustration of the coordinate system with $z$-axis aligned with the BH spin ($J_{\rm BH}$). $\hat{n}$ represents the instantaneous normal to the hot flow (which precesses around the BH spin axis, as the phase $\omega t$ unwinds). The observer direction $\hat{o}$ lies in the $xz$ plane and makes an $i_{\rm jet}$ angle with respect to $z$ axis. The angle $\theta$ represents the viewing angle of the hot flow at phase $\omega t$.}
\label{fig:newgeometry}}
\end{figure}

\begin{figure*}
\centering
\includegraphics[scale=0.7]{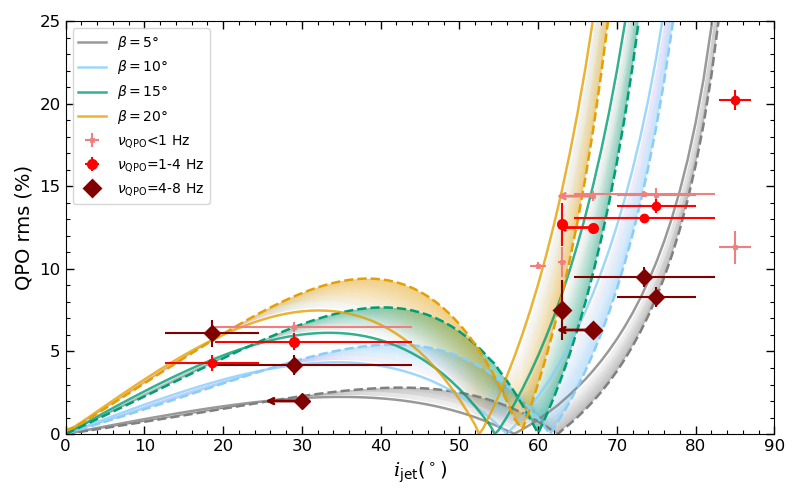}
\caption{QPO fractional rms versus inclination diagram for three different frequency ranges: $\nu_{\rm QPO}<1$~Hz (squares), $1$~Hz$<\nu_{\rm QPO}<4$~Hz (circles) and $4$~Hz$<\nu_{\rm QPO}<8$~Hz (diamonds). Solid lines show the predicted rms from a precessing, Comptonizing slab model for an extended hot flow spanning $3-30R_{\rm S}$, while dashed lines correspond to a narrow ring at $5R_{\rm S}$. Line colors indicate four different misalignment angles $\beta= 5^\circ$ (grey), $10^\circ$ (blue), $15^\circ$ (green) and $20^\circ$ (yellow). The predicted rms for models with different inner and outer radii lie within the shaded region. \label{fig:ltalt}}
\end{figure*}

The hard-state X-ray emission of BH binaries is thought to be produced by Comptonization in a hot medium \citep[see, e.g., the review by][]{dgk07}, that was recently found to be extended in the direction orthogonal to the jet axis \citep{Krawczynski2022,Veledina2023,Ingram2024,mastroserio2025}.
The angular distribution of emission and polarimetric properties originating from such a medium (slab geometry) are anisotropic, hence a precessing slab is expected to produce variations of flux and polarization with the QPO phase \citep[see detailed predictions of this model in][]{veledina_poutanen_ingram_2013,Ingram2015}.

The essence of the precessing Comptonization slab model is the following. 
Locally, the emission is assumed to be produced by the Comptonization processes, with spectral properties being dependent on the angle between the slab normal and the observer direction, $\zeta'$, in the local co-moving frame.
The slab normal is assumed to be rotating (precessing) around a fixed axis, leading to periodic variations of $\zeta'$ as a function of precession phase.
The angular dependence of the emission in the local frame can be approximated with a simple analytical function: $f'(\zeta')=1-0.7\cos\zeta'$ \citep[relevant to the Comptonization processes in a slab,][]{veledina_poutanen_ingram_2013}.
This dependence is modified by  general and special relativistic effects, leading to the altered angular dependence of the emission, which becomes dependent on the position and distance of the emitting element with respect to the BH.
This leads to the dependence of the predicted amplitude and shape of the QPO profiles on the size (outer radius) of the hot flow, i.e., the slab.

{The original model of \citet{veledina_poutanen_ingram_2013} predicts the flux modulation of the precessing flow as a function of both the orbital inclination and the azimuthal angle of the BH spin ($\Phi$ in their notations). 

Since in our new coordinate system the jet is aligned with the z axis, the dependence on $\Phi$ vanishes.
Below we show an explicit expression for the modulation of the viewing angle of the hot flow $\theta$, as a function of the QPO phase $\omega t=2\pi\nu t$, in this frame.
We consider the coordinates where the $z$ axis is aligned with the BH spin and the observer direction ($\hat{o}$) lies in the $xz$ plane (see Fig. \ref{fig:newgeometry}):
\begin{equation}
\hat{o} = (\sin i_{\rm jet}, 0, \cos i_{\rm jet}).
\end{equation}
The instantaneous normal to the hot flow is given by
\begin{equation}
\hat{n} = (\sin\beta\cos\omega t, \sin\beta\sin\omega t, \cos\beta).
\end{equation}

The cosine of the hot flow viewing angle $\theta$ is then given by the scalar product of these two vectors,
\begin{equation}
\cos\theta = \hat{n} \cdot \hat{o} = \sin i_{\rm jet} \sin\beta\cos\omega t + \cos i_{\rm jet} \cos\beta.
\end{equation}
Therefore, the modulation of the viewing angle in this formalism depends only on two parameters: the inclination of the observer with respect to the jet $i_{\rm jet}$ and the misalignment $\beta$.
The QPO profiles additionally depend on the connection between the local angular distribution of emission ($\zeta'$) and the viewing angle ($\theta$), which in turn depends on the proximity of the emitting region to the BH and its extent (hot flow radius).
}

Determining how the radius of the hot flow is related to the QPO frequency  is not straightforward.
A mismatch between the truncation radius inferred from LT precession and spectral modeling is a well-known problem \citep{marcel-neilsen2021,vincentelli2021_1535,nathan2022}. 
{ In this regard, it is important to note that while the original model by \citealt{veledina_poutanen_ingram_2013} was considered in the framework of LT precession, the predictions of the QPO profiles apply to a generic precessing slab.} The model effectively describes LT precession only when linking the truncation radius with the QPO frequency \citep[see e.g.][]{ingram2009}.
Given this,  we compare our results with only two cases:  the QPO signal from (i) a narrow disc annulus located at 5$R_{\rm S}$ ($R_{\rm S}$ is the Schwarzschild radius) and (ii) a flow extended between inner and outer radii of 3 and 30$R_{\rm S}$.
Fig.~\ref{fig:ltalt} shows the data overlaid on top of the model predictions for four different misalignment angles:  $\beta=5^\circ$ (grey), $10^\circ$ (blue), $15^\circ$ (green) and $20^\circ$ (yellow).

We can visualize the predicted curves in 2 regimes:
at low inclinations, the rms as a function of inclination reaches a maximum before falling to zero,  while for high inclinations, the amplitude increases steeply.
Beyond the critical jet inclination of $i_{\rm jet, c} = 90^\circ-\beta$, the assumption of a razor-thin slab leads to self-eclipses and the model predictions become unphysical \citep[as acknowledged in][]{veledina_poutanen_ingram_2013}; however, these cases are beyond the limits shown in Fig.~\ref{fig:ltalt}.
The drop of the rms at inclinations $i_{\rm jet}\sim50-60^\circ$ is caused by the proximity of these inclinations to the peak angle of the Comptonizing slab emission (see fig. 3 and sect. 2.4 of \citealt{veledina_poutanen_ingram_2013}), which leads to an effective ``leakage'' of the QPO rms to the second harmonic, resulting in the harmonic-dominated QPOs (HD-QPOs).
As the flow precesses around the jet axis (assumed to be the black hole spin axis), the observer crosses the emission peak twice per orbit, leading to the dominance of the second harmonic in the X-ray light curve.

The common practice of associating the highest-rms QPO with the fundamental, rather than its harmonic, can lead to a systematic misidentification and complicate testing the model predictions. 
Indications of such HD-QPOs may arise from the detection of ``sub-harmonic'' peaks, which in fact represent the true precession frequency.
Additional constraints may be obtained from multiwavelength data, as the optical and infrared QPOs originating from the hot accretion flow are expected to be dominated by the true fundamental \citep{veledina_poutanen_ingram_2013}. 
A clear candidate for these HD-QPOs can be found in the prototypical system GX~339--4, which shows optical and infrared QPOs in harmonic relation with the X-rays \citep{motch1983,kalamkar2016}
Hence, below we assume the identified rms of the QPOs are associated with the fundamental precession frequency.

Fig.~\ref{fig:ltalt} shows an overall good agreement of the model predictions with the data for misalignment angles $\beta\lesssim20^\circ$.
However, considering that the uncertainties on the disc emission lead to the systematic reduction of the QPO fractional rms, along with the aforementioned simplifications of the model, the inferred misalignment for any particular system can be considered as a lower limit.
The current model cannot predict higher rms for the given misalignment, but many factors can lead to its substantial reduction.
It is interesting to note the specific case of MAXI J1820+070, where a misalignment $\beta>40^\circ$ was measured \citep{Poutanen2022}.
In Fig.~\ref{fig:ltalt}, the rms of its QPOs at ${<}1$~Hz  is consistent with the yellow solid line, i.e. an extended flow with misalignment $\beta=20^\circ$. This case shows that the reduction of the predicted rms in this simplified model can be severe.

Finally, our findings can also be compared to the predictions of the binary evolution computations of \citet{fragos2010}, which find over 67\% of binaries to have misalignment $\beta\lesssim10^\circ$. 
We find that 5 out of 10 systems in our sample lie above the lines of $\beta=10^\circ$ (blue), hence our results are marginally consistent with the predictions of the \citet{fragos2010} binary evolution model.

\section{Conclusions}

We performed a systematic analysis of the QPO rms evolution across the \textit{RXTE} and \textit{HXMT} archives to quantify the dependence of Type C QPOs on inclination for different frequency ranges up to 8 Hz.  Our analysis significantly improves previous studies for three main reasons. First, we used radio measurements to infer the jet inclination. Second, to our knowledge, this is the first study comparing QPOs using both satellites. Last, we used only the fundamental component, allowing a direct comparison with QPO models. Interestingly, our analysis revealed  a systematic difference between the amplitude of QPOs during the rise and decay phases of the outbursts. This suggests that the hot flow (or corona) has different properties during the rising and decaying phases of the outburst \citep[as argued in, e.g.,][]{barnier2022}, although this difference does not seem to impact the polarization \citep{Podgorny2024}.

This database provides a simple yet powerful benchmark against which viable QPO models must be tested. The observed rms-inclination relation is difficult to reconcile with models in which the QPO arises solely from an intrinsically isotropic luminosity oscillation. Instead, it favours scenarios in which the modulation is associated with a structural,  geometry-dependent variation of the X-ray emitting region (e.g., a precessing inner hot flow or corona, possibly combined with relativistic and anisotropic emission effects), which naturally produce stronger modulations at high inclinations. By modelling the expected rms–inclination dependence within a simplified precessing-slab framework, we could constrain for the first time the minimum spin-orbit misalignment required to reproduce the observations. Under the assumptions of this model, the observed rms-inclination relation is consistent with a spin-orbit misalignment of at least $\approx$10–15$^\circ$ for most sources. Given the simplifying assumptions of the model and the possible dilution from disc emission, this value should be regarded as a lower limit rather than a precise measurement. In the future, detailed modeling of the absolute rms spectrum evolution of individual sources will enable much tighter constraints on the spin-orbit misalignment. In combination with measurements of systemic peculiar velocities—an independent tracer of natal kicks—such constraints may provide a promising avenue to discriminating between different binary-evolution pathways \citep{zhao2023}.

Finally, we highlight how the robustness of this study would greatly benefit from the inclusion of systems with jet inclinations around $i_{\rm jet} \approx 40-60^\circ$. To our knowledge, no confirmed black hole systems currently populate this inclination range \citep[see also Table~1 in][]{2024A&A...681A..49P}.  In the precessing Comptonizing slab scenario considered here, in such an inclination range, these systems are expected to exhibit a lower amplitude of the first harmonic, with a QPO dominated by the second harmonic. This may be the case for GX 339--4, where optical and infrared QPOs appear to coincide with the X-ray sub-harmonic \citep{motch1983,kalamkar2016}. If so, we would expect a jet inclination of $i_{\rm jet} \approx 40$--$60^\circ$.  Expanding the sample in the future will allow firmer constraints on the inner geometry of the accretion flow and provide a decisive test of the trend reported in the present study.

\section*{Acknowledgements}

{The authors are thankful to the anonymous referee for their constructive comments which significantly improved the paper.}

 FMV acknowledges support by the European Union’s Horizon Europe research and innovation programme with the Marie Sk\l{}odowska-Curie grant agreement No. 101149685 and the Visitor and Mobility program of the Finnish Centre for Astronomy with ESO (FINCA).
FMV and DA acknowledge support from the Science and Technology Facilities Council grant ST/V001000/1. FMV also thanks Poshak Gandhi and Mariano Mendez for the insightful discussions.

AV and GM acknowledge support from the Research Council of Finland grants 355672 and 372881. GM acknowledges support from the Polish National Science Center grant 2023/48/Q/ST9/00138. 
Nordita is supported in part by NordForsk. 

The authors are thankful for the team meetings at the International Space Science Institute (ISSI) for fruitful discussions. In particular this research was supported by ISSI in Bern, through the International Team projects 'Looking at the Disc-Jet Coupling from Different Angles: Inclination Dependence of Black-Hole Accretion Observables' (\#18-440) and  'What are the spins of stellar-mass black holes?’ (ISSI Team project \#25-660) 

Y.C. acknowledges support from the grant RYC2021-032718-I, financed by MCIN/AEI/10.13039/501100011033 and the European Union NextGenerationEU/PRTR.

%%%%%%%%%%%%%%%%%%%%%%%%%%%%%%%%%%%%%%%%%%%%%%%%%%
\section*{Data Availability}

RXTE raw fits files are available on the \textsc{heasarc} archive, while the HMXT data are available at this website: \url{http://archive.hxmt.cn/proposal}. The average rms in the 4 frequency bands are reported in Tab. \ref{taB:tab}. An updated version of the data reported in Table 1  will be available on the following website: \url{https://github.com/fvince90/QPO-rms-vs-Jet-Inclination/tree/main} The fit of the individual power spectra is available on request.



%\begin{thebibliography}{99}
%\end{thebibliography}

%%%%%%%%%%%%%%%%%%%%%%%%%%%%%%%%%%%%%%%%%%%%%%%%%%

%%%%%%%%%%%%%%%%% APPENDICES %%%%%%%%%%%%%%%%%%%%%

\appendix

\section{Formula for the HXMT rms}

Estimating the power $P$ of the QPO by combining to adjacent energy bands is equivalent to computing the variance of the sum of two random variables $X$ and $Y$. This is
\begin{equation}
\sigma^2(X+Y)= \sigma^2_X+\sigma^2_Y+2\cdot\sigma_{X,Y}
\label{egen}
\end{equation}

Where $\sigma^2_i$ represents the variance  of a random variabile $i$ and $\sigma_{X,Y}$  the covariance between the two variables $X$ and $Y$.  The covariance in the Fourier domain is equivalent to the amplitude of the cross-spectrum $C_{XY}$. This can be easily computed from the intrinsic coherence $\gamma^2$, which is \citep{vaughannowak1997,uttley2014}:

\begin{equation}
\begin{split}
\gamma^2=|C_{XY}|^2/P_XP_Y
\end{split}
\end{equation}

Where $P_X$ and $P_Y$ are the power spectra in the two bands (i.e. $\sigma^2$). As demonstrated by several studies, the coherence of the QPO in the energy bands between 2-25 keV  can be safely considered as unity \citep{Bollemeijer2025}. This means that the previous equation can be arranged in the following way:
\begin{equation}
\begin{split}
C_{XY}=\sqrt{P_XP_Y}
\label{cov}
\end{split}
\end{equation}

Thus, by substituting eq. \ref{cov} in eq. \ref{egen}, we obtain:

\begin{equation}
\label{eq:stoold}
\begin{split}
\sigma^2(X+Y)= \sigma^2_X+\sigma^2_Y+2\cdot\sqrt{\sigma_X^2\sigma_Y^2}
\end{split}
\end{equation}

These equations represent the new variance in \textit{absolute} units. We can now express it in relative units (i.e. fractional rms units) as $\sigma^2_X=CR_X^2\cdot rms_X^2$, where $CR$ is the count rate and $rms^2$ is the fractional squared rms in the band $X$. By using this relation, we can rearrange eq. \ref{eq:stoold} and obtain

\begin{equation}
\label{eq:sto}
\begin{split}
rms^2(X+Y) \cdot (CR_X+CR_Y)^2= rms^2_X \cdot CR_X^2+rms^2_Y \cdot CR_Y^2\\+2\cdot\sqrt{rms_X^2 \cdot CR_X^2 \cdot rms_Y^2 \cdot CR_Y^2}
\end{split}
\end{equation}
which corresponds to eq. \ref{eq:rmseq}%%%%%%%%%%%%%%%%%%
% Don't change these lines

\bsp	% typesetting comment
\label{lastpage}
\end{document}